%% file: lists.tex
\documentclass{llncs}
\usepackage{graphicx}
\usepackage{subfigure}
\usepackage{url}
\usepackage{common}

\title{Identifying Topical Twitter Communities \\via User List Aggregation}
\author{Derek Greene, Derek O'Callaghan, P\'{a}draig Cunningham }
\institute{School of Computer Science \& Informatics, University College Dublin \\
\email{\{derek.greene,derek.ocallaghan,padraig.cunningham\}@ucd.ie} }

\begin{document} 
\maketitle
\begin{abstract}
\input{abstract}
\end{abstract}


\input{intro}
\input{related}

\input{methods}
\input{eval}

\input{conc}

\vspace{3 mm}\noindent\emph{Acknowledgements.} 
This research was supported by Science Foundation Ireland Grant 08/SRC/I1407 (Clique: Graph and Network Analysis Cluster).


\bibliographystyle{splncs}
\bibliography{userlists}

\end{document}

%% file: abstract.tex

A particular challenge in the area of social media analysis is how to find communities within a larger network of social interactions. Here a community may be a group of microblogging users who post content on a coherent topic, or who are associated with a specific event or news story.
Twitter provides the ability to curate users into lists, corresponding to meaningful topics or themes. Here we describe an approach for crowdsourcing the list building efforts of many different Twitter users, in order to identify topical communities. This approach involves the use of ensemble community finding to produce stable groupings of user lists, and by extension, individual Twitter users. We examine this approach in the context of a case study surrounding the detection of communities on Twitter relating to the London 2012 Olympics.

%% file: intro.tex
\section{Introduction}

A wide variety of community finding techniques have been proposed in the literature, with recent research focusing on the challenge of identifying overlapping communities \cite{fortunato10review}. In the case of microblogging data, researchers have been interested in the identification of communities of users on Twitter, who produce tweets on a common topic, who belong to the same demographic, or who share a common ideological viewpoint \cite{java07twitter}. These approaches have generally relied on explicit views of the Twitter network, such as follower relations or retweets.

\begin{figure}[!t]
	\begin{center}
	\includegraphics[width=0.82\linewidth]{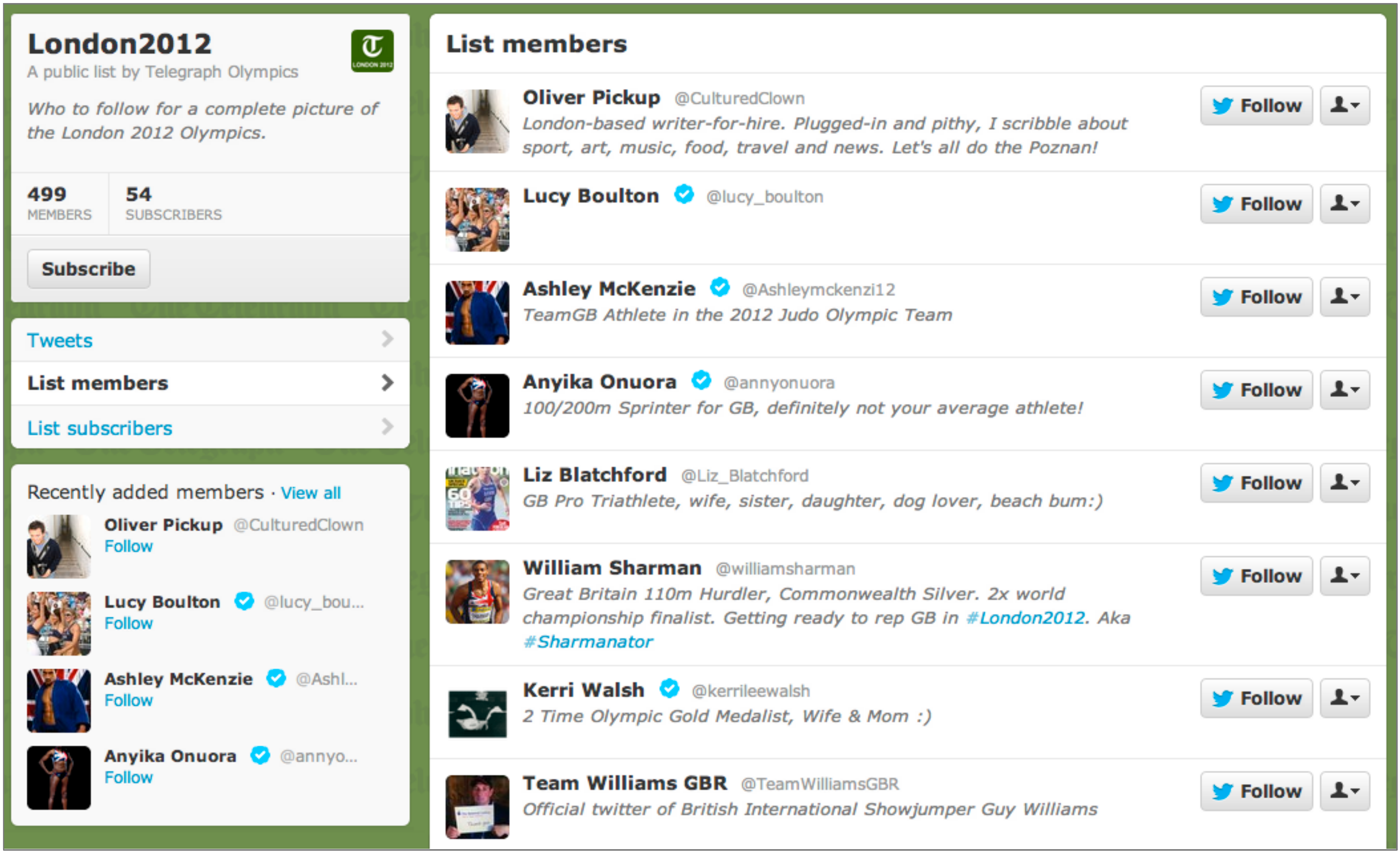}
	\end{center}
	\vskip -1.0em
	\caption{User list, curated by The Telegraph, covering athletes, journalists, and organisations involved in the London 2012 Summer Olympics.}
	\label{fig:list}
\end{figure}

Twitter users can organise the accounts that they follow into Twitter \emph{user lists}, as shown in \reffig{fig:list}. 
These lists are used in a variety of ways. In some cases they may correspond to personal lists of a given user's friends and families, but frequently lists are employed to group together Twitter accounts based on a common topic or theme. In this way, every Twitter user can effectively become a community curator. Notably, journalists from news organisations such as The Telegraph and Storyful curate lists relevant to a given news story or event, as a means of monitoring breaking news. Recently,  Kim \etal and Garc\'{i}a-Silva \etal \cite{kim10lists,garcia12lists} both discussed the potential of user lists to provide latent annotations for Twitter user profiles, while Wu \etal \cite{wu11says} suggested user lists as a means of harnessing the ``wisdom of the crowds'' on Twitter.

Our primary goal here is to demonstrate that topical communities can be identified by harnessing the  ``crowd-sourced'' list building efforts of a large base of Twitter users. In \refsec{sec:methods}, we show that this can be done by constructing a graph based on the similarity of user list memberships, and then using an \emph{ensemble community finding} approach to find robust, overlapping groups of lists within this graph, from which user communities can be derived. We use stability information derived from the ensemble as a proxy for the reliability of the communities. In \refsec{sec:eval}, we evaluate the proposed techniques on a case study  relating to coverage of the London 2012 Olympics on Twitter.

%% file: related.tex
\section{Related Work}
\label{sec:related}

\subsection{Microblogging Analysis}

Many researchers have become interested in exploring the network structure within the Twitter network, given the potential for Twitter to facilitate the rapid spread of information. Java \etal \cite{java07twitter} provided an initial analysis of the early growth of the network, and also performed a small-scale evaluation that indicated the presence of distinct Twitter user communities, where the members share common interests as reflected by the terms appearing in their tweets. 
Kwak \etal \cite{kwak10twitter} performed an evaluation based on a sample of 41.7 million users and 106 million tweets from a network mining perspective. The authors studied aspects such as: identifying influential users, information diffusion, and trending topics. 
Shamma \etal \cite{shamma09debates} performed an analysis on microblogging activity during the 2008 US presidential debates. The authors demonstrated that frequent terms reflected the topics being discussed, but the use of informal vocabulary complicated topic identification.

Typically researchers have focused either on Twitter users from the perspective of the content that they produce, or in terms of explicit network representations based on follower relations or retweeting activity \cite{kwak10twitter}. However, preliminary work by Kim \etal \cite{kim10lists} suggested that latent groups and relations in Twitter data could be extracted by examining user list data. 
Wu \etal \cite{wu11says} suggested that user list memberships could be used to organise users into a pre-defined set of categories: celebrities, media, organisations, and blogs. 
Having classified the users, important or ``elite'' users within each category were identified based on the number of lists to which each user was assigned.
Garc\'{i}a-Silva \etal \cite{garcia12lists} described approaches for extracting semantic relations from user lists, by constructing relations between co-occurring keywords taken from list names. 
In the context of list curation,  Greene \etal \cite{greene11curation} showed that \emph{co-listed} links could be used as one potential way to recommend users who may belong to the ``community'' surrounding a breaking news story.

\subsection{Community Finding}
\label{sec:related_community}
Many different algorithms have been proposed to identify communities in graphs, based on different combinations of objective functions and search strategies \cite{fortunato10review}. Recently, considerable focus has been given to identifying communities in highly-overlapping, weighted networks. A widely-employed algorithm in this area is OSLOM (Order Statistics Local Optimization Method), introduced by Lancichinetti \etal \cite{lancichinetti11oslom}.
Kwak \etal \cite{kwak09consistent} observed that many community detection algorithms can produce inconsistent results, due to stochastic elements in their optimisation process. Lancichinetti \& Fortunato \cite{lanc12consensus} demonstrated that this also applied to OSLOM, and proposed an ensemble approach to generate stable results out of a set of multiple partitions.

In the more general cluster analysis literature, \emph{ensemble clustering} methods have been previously developed to address such issues. These methods typically involve generating a diverse set of ``base clusterings'', which are then aggregated to produce a consensus solution \cite{fred01consistent,strehl02cluster,topchy03combination}. The most popular aggregation strategy has been to use information derived from different clusterings to determine the level of association between each pair of items in a dataset \cite{fred01consistent,strehl02cluster}. This strategy was motivated by the observation that pairwise co-assignments, averaged over a sufficiently large number of clusterings, may be used to induce a new, more robust measure of similarity on the data.

%% file: methods.tex
\section{Methods}
\label{sec:methods}

In this section, we introduce an approach that aggregates user list information to generate communities. Firstly, we describe the construction of a graph representation of user lists, based on their membership overlaps. Then in \refsec{sec:combine} we describe an ensemble approach to identify overlapping groups of user lists. The stability of these groups is assessed as described in \refsec{sec:stability}, and the selection of  community labels is discussed in \refsec{sec:tags}. Finally, the derivation of corresponding communities for individual users is discussed in \refsec{sec:user}.

\subsection{User List Graph Construction}
\label{sec:graph}

We construct a graph $G$ of $l$ nodes, where each node represents a distinct Twitter user list $L_{x}$. A weighted edge exists between a pair of lists if they share users in common. Rather than using the raw  intersection size between a pair, we make allowance for the significance of the intersection size relative to the size of the two lists, and the total number of users assigned to lists $n$. For a pair $(L_{x},L_{y})$, we compute a $p$-value to indicate the significance of the probability of observing at least $\abs{L_{x} \cap L_{y}}$ users from $L_{x}$ within another list of size $\abs{L_{y}}$:
\begin{equation}
PV(L_{x},L_{y}) = 1 - \sum_{j=0}^{\abs{L_{x} \cap L_{y}}-1} 
\frac{\binom{\abs{L_{x}}}{j}\binom{n-\abs{L_{x}}}{\abs{L_{y}}-j}}{\binom{n}{\abs{L_{y}}}}
\label{eqn:pvalue}
\end{equation}
To improve interpretability, we compute the associated log $p$-value:
\begin{equation}
LPV(L_{x},L_{y}) = -\textrm{log}\left( PV(L_{x},L_{y}) \right)
\label{eqn:lpv}
\end{equation}
where a larger value is more significant. We consider \reft{eqn:lpv} as a measure of the similarity between a pair of user lists, corrected for chance. To further increase the sparseness of the graph, we remove edges with weights $LPV < \rho$ for a weight  threshold $\rho$. Increasing the value of $\rho$ will result in an increasingly sparse graph.


\subsection{Combining Overlapping Communities}
\label{sec:combine}

We will naturally expect that different topical communities will potentially overlap with one another. To identify communities of lists, we apply the OSLOM algorithm which has been shown to out-perform other community finding approaches \cite{lancichinetti11oslom}. However, as noted in \cite{lanc12consensus}, OSLOM can produce unstable results.

Following the CSPA ensemble aggregation approach \cite{strehl02cluster}, and the method for combining network partitions \cite{lanc12consensus},  we now describe an approach for generating and combining an ensemble of overlapping community sets. Given an initial user list graph $G$, we construct a symmetric $l \times l$ \emph{consensus matrix} $\m{M}$. For the purpose of generating a collection of $r$ \emph{base community sets}, we apply the OSLOM algorithm \cite{lancichinetti11oslom} using a different initial random seed for each run. Motivated by the notion an ensemble of weak clusterings \cite{topchy03combination}, 
we use the ``fast'' configuration of OSLOM, which uses a minimal number of optimisation iterations.  

\begin{figure}[!t]
\centering
\includegraphics[width=0.75\linewidth]{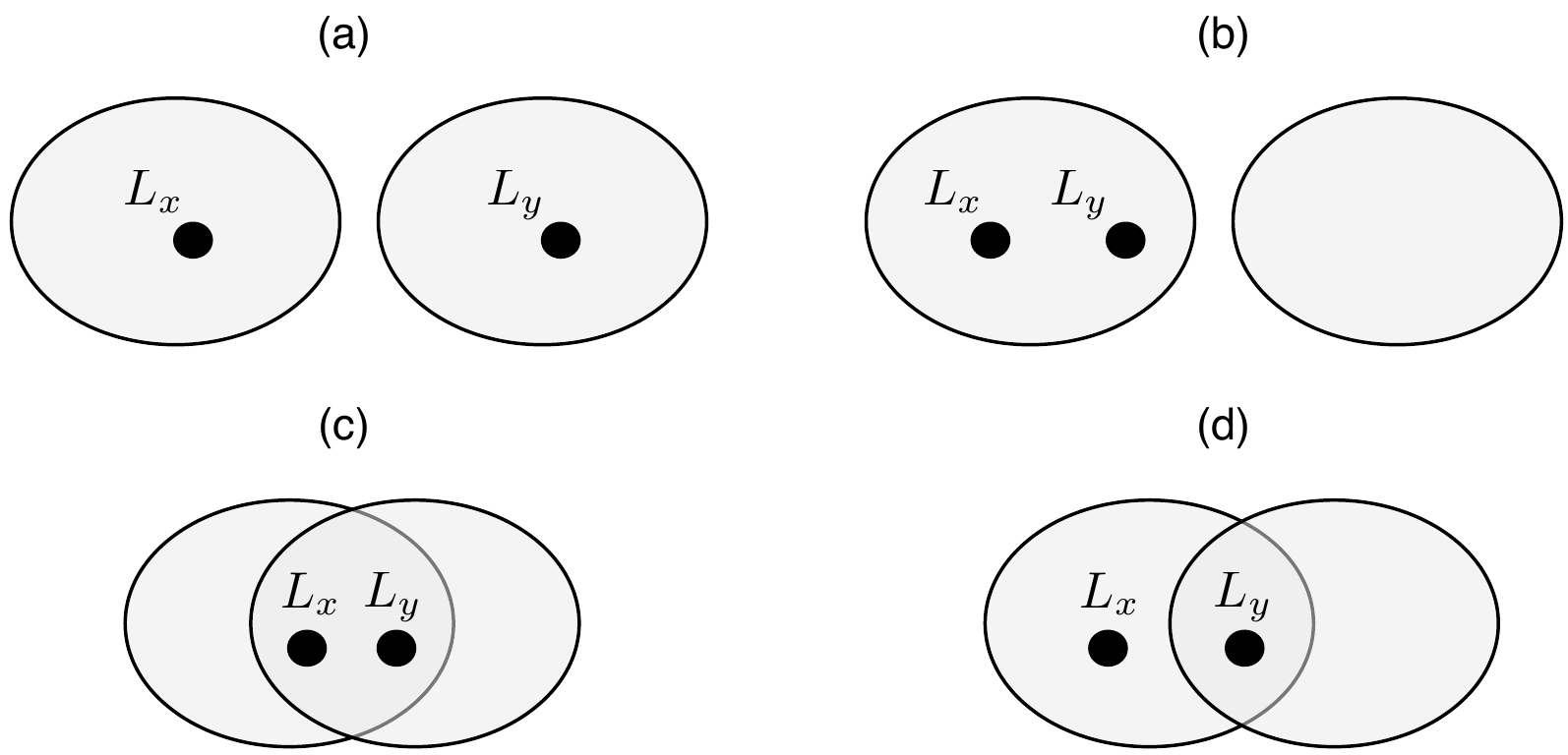}
\vskip -0.55em
\caption{Four different cases of computing the Jaccard similarity between the sets of community labels assigned to two nodes $L_{x}$ and $L_{y}$. The Jaccard scores respectively are: (a) 0.0, (b) 1.0, (c) 1.0, (d) 0.5. }
\label{fig:jaccard}
\end{figure}

After generating a base community set, for each unique pair of nodes $(L_{x},L_{y})$ in network, we compute the Jaccard similarity between the sets of community labels assigned to those nodes by OSLOM. If the pair are not both co-assigned to any community, the score is 0. If the pair are present in all communities together, the score is 1. However, unlike the binary approach of \cite{lanc12consensus}, if the pair are present in some but not all communities together, the Jaccard score will reflect this. See \reffig{fig:jaccard} for  examples. In the case of non-overlapping partitions, the score will reduce to the binary scoring used in \cite{lanc12consensus}. After computing all Jaccard scores, we increment the corresponding matrix entries in $\m{M}$. Note that, by definition, singleton communities are ignored during this aggregation process.

Once all $r$ base community sets have been generated, $\m{M}$ is normalised by $1/r$ to give a matrix with entries $\in [0,1]$. 
To find the \emph{consensus communities}, we follow a similar approach to that used by \cite{lanc12consensus}. We construct a new undirected weighted graph such that, for every unique pair of nodes $(L_{x},L_{y})$, we create an edge with weight $M_{xy}$ if $M_{xy} \geq \tau$. The threshold parameter $\tau \in [0,1]$ controls the sparsity of the graph. We then apply OSLOM to this graph for a large number of iterations ($\approx 50$) to produce a final grouping of the user lists in $G$.

\subsection{Evaluating Community Stability}
\label{sec:stability}

When applying community detection, often we may wish to examine the most reliable or robust communities with the strongest signals in the network. Here, we rank the consensus communities generated as described in \refsec{sec:combine}, based on the cohesion of their members with respect to the consensus matrix $\m{M}$. A more stable consensus community will consist of user lists which were frequently co-assigned to one or more communities across all $r$ base community sets.

For a given consensus community $C$ of size $c$, we compute the mean of the values $M_{xy}$ for all unique pairs $(L_{x},L_{y})$ assigned to $C$; this value has the range $\in [0,1]$. We then compute the mean expected value for a community of size $c$ as follows: randomly select $c$ unique nodes from $G$, and compute their mean pairwise score from the corresponding entries in $\m{M}$. This process is repeated over a large number of randomised runs, yielding an approximation of the expected stability value. We then employ the widely-used adjustment technique introduced by \cite{hubert85compare} to correct stability for chance agreement: 
\begin{equation}
\textrm{CorrectedStability}(C) = \frac{\textrm{Stability}(C) - \textrm{ExpectedStability}(C)}{1 - \textrm{ExpectedStability}(C)}
\label{eqn:stability}
\end{equation}
A value close to 1 indicates a highly-stable community, while a value closer to 0 is indicative of a weak community that appeared intermittently over the $r$ runs. We rank all consensus communities based on their values for \reft{eqn:stability}.

\subsection{Selecting Community Labels}
\label{sec:tags}

To summarise the content of a consensus community, we aggregate the meta-information associated with all lists assigned to that community. Specifically, we construct a bag-of-words model, where each user list is represented by unigrams and bigrams tokenised from the list's name and description. Single stop-words are removed, and terms are weighted using log-based TF-IDF. For each community, we then compute the centroid vector corresponding to the mean vector of all lists assigned to that community. To generate descriptive labels for the community, we subtract the mean vector of {\bf{all user list vectors}} from the community centroid vector, and rank terms in descending order based on the resulting weights. The top ranked terms are used as community labels.

\subsection{Deriving User Memberships}
\label{sec:user}

The consensus communities generated using the method proposed in \refsec{sec:combine} can potentially provide us with an insight into the overall topics in a Twitter corpus. However, it will often be useful to assign community memberships to individual users. We can readily produce this by using the list groupings in conjunction with the original user list membership information.

For a given consensus community $C$ of size $c$, we examine the memberships of all lists in the community.
We consider the assignment of a user $u_{i}$ to each $L_{x} \in C$ as being a vote with weight $1/c$ for $u_{i}$ belonging to the overall community. The total membership weight for $u_{i}$ is therefore given by the fraction of lists $L_{x} \in C$ containing $u_{i}$. Membership weights for all communities are computed in this way.  We can also rank the importance of users in a given community by sorting users by weight in descending order. To produce a final set of user communities, we only include a user in a community for which the user has a membership weight $\geq \mu$, based on a membership threshold $\mu \in [0,1]$.

%% file: eval.tex
\section{Evaluation}
\label{sec:eval}

\subsection{Data Collection}
To evaluate the proposed community finding methods, we constructed a dataset based on a list of 499 users curated by The Telegraph, which covers athletes, journalists, and organisations involved in the London 2012 Summer Olympics\footnote{\url{http://twitter.com/#!/Telegraph2012/london2012}}. Initially, for each user we retrieved up to their 200 most recent user list assignments. From this initial pool of lists, we then retrieved list memberships for 10,000 randomly selected lists of size $\geq 5$ and containing at least 2 core list users. This yielded a dataset containing a total of 44,484 individual list membership records, where the average number of lists per user was 89. The most frequently-listed user was \emph{@andy\_murray}, assigned to 1,931 different lists.

\subsection{Community Detection}
Using the approach described in \refsec{sec:graph}, we constructed a user list graph based on membership information for the 499 users. To limit the density of the graph, we use a weight significance threshold of $\rho=6$ (\ie user list overlaps are considered as significant for $LP \leq 1^{-6}$). This resulted in a graph containing 4,948 nodes representing user lists, with 749,062 weighted edges between them. 

To generate an ensemble of base community sets, we apply OSLOM as described in \refsec{sec:combine} for 100 random runs, selecting the lowest level of the hierarchy as the solution for each run. The average number of non-singleton communities in each run was 157. Combining the base community sets yielded a consensus matrix containing $\approx 11.8$m non-zero values. We examined a range of threshold values $\tau \in [0.1,0.5]$, and selected a threshold $\tau=0.2$ to generate consensus communities in order to maximise coverage over user lists, while also reducing the density of the consensus graph. Applying OSLOM to the sparse graph of $\approx 1.5$m values produced a total of 94 consensus communities, considerably lower than the average base community count. Finally, user communities were derived using a low membership threshold $\mu=0.1$ to maximise the number of core users assigned to communities. In total, 416 core users were assigned to at least one community, representing 83\% of the overall total. Of these, 362 users were assigned to multiple communities. 

\begin{table}[!b]
\caption{Top 15 user list communities, arranged in descending order by stability score.}
\begin{center}
\vskip -0.6em
\scriptsize{
\begin{tabular}{|c|cc|p{4.45cm}|p{5.08cm}|}\hline
\bf Score & \bf Lists & \bf Users & \bf Top Labels & \bf Top Users \\ \hline
1.00 & 17 & 14 & badminton, badminton players,\;\;\;\;\;\;\;\;\; badders & @Jennywallwork, @Nath\_Robertson,\;\;\;\;\;\;\;\; @ChrisAdcock1 \\ \hline
1.00 & 5 & 5 & bmx, bmx racing, bmx atl\~{e}ti & @ShanazeReade, @liamPHILLIPS65,\;\;\;\;\;\;\; @bloomy181 \\ \hline
1.00 & 32 & 11 & sailing, sailors, olympic & @SkandiaTeamGBR, @AinslieBen,\;\;\;\;\;\;\;\;\;\;\; @matchracegirls \\ \hline
1.00 & 19 & 24 & fencing, fencers, individuele schermers & @britishfencing, @CBennettGBR,\;\;\;\;\;\;\;\;\;\;\;\; @LaurenceHalsted \\ \hline
1.00 & 6 & 5 & triathlon, machines, swim run & @AliBrownleetri, @jodieswallow,\;\;\;\;\;\;\;\;\;\;\;\;\;\; @MarkCavendish \\ \hline
1.00 & 5 & 21 & scots, red sky, 2014 & @mj88live, @RobbieRenwick,\;\;\;\;\;\;\;\;\;\;\;\;\;\;\;\;\;\;\; @Euan\_Burton \\ \hline
1.00 & 22 & 4 & wielrennen, ciclismo, cycling & @GeraintThomas86, @UCI\_cycling,\;\;\;\;\;\;\;\;\; @MarkCavendish \\ \hline
0.98 & 5 & 7 & track, field, track field & @allysonfelix, @TysonLGay, @tiffofili \\ \hline
0.98 & 48 & 19 & rowing, rowers, gb rowing & @andrewthodge, @ZacPurchase,\;\;\;\;\;\;\;\;\;\;\;\;\;\; @MarkHunterGB \\ \hline
0.97 & 14 & 22 & diving, tuffi, olympic diving & @PeterWaterfield, @matthew\_mitcham,\;\;\; @toniacouch \\ \hline
0.96 & 36 & 44 & hockey, hockey players, field hockey & @AlexDanson15, @RichM6, @jfair25 \\ \hline
0.96 & 12 & 21 & canoe, canoeing, canoe slalom & @GBcanoeing, @PlanetCanoe,\;\;\;\;\;\;\;\;\;\;\;\;\;\;\;\;\; @edmckeever \\ \hline
0.93 & 5 & 5 & actors athletes, internet stars,\;\;\;\;\;\;\;\;\;\;\; athletes tmz & @usainbolt, @ShawnJohnson,\;\;\;\;\;\;\;\;\;\;\;\;\;\;\;\;\;\; @MichaelPhelps \\ \hline
0.92 & 27 & 13 & judo, judo clubs, judo related & @BritishJudo, @USAJudo, @IntJudoFed\; \\ \hline
0.87 & 6 & 7 & runners, hardlopen, runners world & @Mo\_Farah, @paulajradcliffe,\;\;\;\;\;\;\;\;\;\;\;\;\;\;\;\;\;\; @KenenisaBekele \\ \hline
\end{tabular}
}
\end{center}
\vskip -2.8em
\label{tab:communities}
\end{table}

\reftab{tab:communities} shows the top 15 communities, arranged in descending order by their stability score, as defined in \reft{eqn:stability}. The table shows the size of each community (in terms of both number of lists and users assigned), the top text labels selected for each community, and the three highest-weighted users. We observe that the most stable communities generally correspond to communities of users involved in specific, ``niche'' sports (\eg badminton, BMX racing, fencing). In these cases, the top-weighted users correspond to either British Olympic athletes competing in these sports, or accounts of the official British organisations for these sports.  
Interestingly, we also see some unexpected communities with high stability - a community around the Glasgow 2014 Commonwealth Games, and a community of celebrities which includes ``elite'' user accounts with hundreds of thousands of followers (\eg \emph{@usainbolt}, \emph{@MichaelPhelps}). 
As stability decreases, we observed that communities become less homogeneous, covering highly-popular sports (\eg football, basketball), or containing users and lists related to several sports. This suggests that the proposed stability provides a useful measure of the homogeneity of topical content for Twitter communities.

Many of the top labels selected for communities are multi-lingual. For instance, the label for the ``cycling'' community in \reftab{tab:communities} contains terms in Dutch, Italian, and English. Unlike in certain textual analyses of tweets, the use of list membership information allows us to identify groups of  users in a language-agnostic manner.

\subsection{External Validation}

To validate the consensus user communities that were identified by aggregating list information, we use a set of fine-grained Olympics lists also produced by The Telegraph\footnote{\url{http://twitter.com/#!/Telegraph2012/lists}}, consisting of Twitter users associated with individual sports (\eg ``archery'', ``equestrianism''). This provided us with 18 external ``ground truth'' categories, covering 423 of the 499 users in the dataset. 

We computed \emph{precision}, \emph{recall}, and \emph{F1} scores for all communities, and subsequently matched  categories to communities based on precision. 
\begin{table}[!h]
\caption{Validation scores achieved relative to 18 ``ground truth'' categories.}
\begin{center}
\vskip -0.5em
\scriptsize{
\begin{tabular}{|l|c|ccc|}\hline
\bf Category Name & \bf Category Size & \bf \;\;Precision\;\; & \bf \;\;Recall\;\; & \bf \;\;\;\;\;\;F1\;\;\;\;\;\; \\ \hline
\it judo & 20 & 1.00 & 0.65 & 0.79 \\ 
\it basketball & 26 & 1.00 & 0.50 & 0.67 \\ 
\it rowing & 44 & 1.00 & 0.43 & 0.60 \\ 
\it athletics & 50 & 1.00 & 0.22 & 0.36 \\ 
\it cycling & 28 & 1.00 & 0.14 & 0.25 \\ 
\it hockey & 47 & 0.98 & 0.91 & 0.95 \\ 
\it diving & 23 & 0.95 & 0.91 & 0.93 \\ 
\it equestrianism & 18 & 0.94 & 0.83 & 0.88 \\ 
\it fencing & 23 & 0.88 & 0.91 & 0.89 \\ 
\it sailing & 16 & 0.82 & 0.56 & 0.67 \\ 
\it gymnastics & 24 & 0.77 & 0.42 & 0.54 \\ 
\it canoeing & 22 & 0.76 & 0.73 & 0.74 \\ 
\it beach-volleyball & 12 & 0.55 & 1.00 & 0.71 \\ 
\it boxing & 22 & 0.55 & 0.55 & 0.55 \\ 
\it swimming-syncrho\;\;\;\;\; & 16 & 0.33 & 0.13 & 0.18 \\ 
\it weightlifting & 6 & 0.20 & 0.17 & 0.18 \\ 
\it archery & 17 & 0.20 & 0.06 & 0.09 \\ 
\it waterpolo & 22 & 0.05 & 0.05 & 0.05 \\ \hline
\end{tabular}
}
\end{center}
\vskip -1.0em
\label{tab:external}
\end{table}
\reftab{tab:external} shows the resulting scores for all categories, arranged in descending order by precision. Communities produced by user list aggregation allowed us to identify eight categories with precision $\geq 0.9$, while generally maintaining high recall.
Only in the case of four categories did the proposed approach lead to precision and recall scores of both $\leq 0.5$. Subsequent examination of the data suggests that list information was relatively sparse for these categories, and that the users were generally assigned to more generic lists (\eg ``aquatics'' for ``waterpolo''). In the case of the ``cycling'' category, which received high precision but low recall, we note that many of the lists we collected related to road racing, whereas many of the users in the category are associated with Olympic velodrome racing.

%% file: conc.tex
\section{Conclusions}

In this paper, we have presented initial work on the idea of identifying topical communities on Twitter by aggregating the ``wisdom of the crowds'', as encoded in the form of user lists. We show that this information can be mined to detect and label coherent overlapping clusters of both lists and users. 

While the evaluation in this paper used a fixed network of users, a similar approach could be applied to identify topical sub-communities around trending terms or hashtags, by compiling a network of users frequently mentioning these terms. Also, in some cases, not all users will have been assigned to any user lists.  We suggest that a classification process, using an alternative network view (\eg follower links) could be used to assign such users to communities.